\documentclass{elsart}

\usepackage{amsmath}
\usepackage{amssymb}
\usepackage{amsfonts}
\usepackage{array}
\usepackage{cite}  
\usepackage{graphicx}
\usepackage{psfrag}
\usepackage{lineno}
\usepackage{url}
\usepackage[dvips,bookmarks=true,pdfborder={0 0 0}, colorlinks, citecolor=black, filecolor=black, linkcolor=black, urlcolor=black]{hyperref} 

\begin{document}

\begin{frontmatter}

\journal{arXiv}

\title{AICA: a New Pair Force Evaluation Method for Parallel Molecular Dynamics in Arbitrary Geometries}

\author{Graham B. Macpherson\corauthref{cor}} and
\ead{graham.macpherson@strath.ac.uk}
\author{Jason M. Reese}
\ead{jason.reese@strath.ac.uk}

\corauth[cor]{Corresponding author.}

\address[strath]{Department of Mechanical Engineering, University of Strathclyde,\\Glasgow, G1 1XJ, UK}

\begin{abstract}

A new algorithm for calculating intermolecular pair forces in Molecular Dynamics (MD) simulations on a distributed parallel computer is presented.  The Arbitrary Interacting Cells Algorithm (AICA) is designed to operate on geometrical domains defined by an unstructured, arbitrary polyhedral mesh, which has been spatially decomposed into irregular portions for parallelisation.  It is intended for nano scale fluid mechanics simulation by MD in complex geometries, and to provide the MD component of a hybrid MD/continuum simulation.  AICA has been implemented in the open-source computational toolbox OpenFOAM, and verified against a published MD code.

\end{abstract}

\end{frontmatter}

\section{Introduction and Motivation}

\subsection{MD Simulation in Arbitrary Geometries}

Simulations of nano scale liquid systems can provide insight into
many naturally-occurring phenomena, such as the action of proteins
that mediate water transport across biological cell
membranes~\cite{NL_Agre_Aquaporin}.  They may also facilitate the
design of future nano devices and materials (e.g.\ high-throughput,
highly selective filters or lab-on-a-chip components). The dynamics
of these very small systems are dominated by surface interactions,
due to their large surface area to volume ratios.  However, these
surface effects are often too complex and material-dependent to be
treated by simple phenomenological parameters
\cite{NL_Koplik_Banavar} or by adding `equivalent' fluxes at the
boundary \cite{MD_Brenner_Ganesan}. Direct simulation of the fluid
using molecular dynamics (MD) presents an opportunity to model these
phenomena with minimal simplifying assumptions.

\begin{figure}[htb]
    \centering
    \includegraphics[width=8cm]{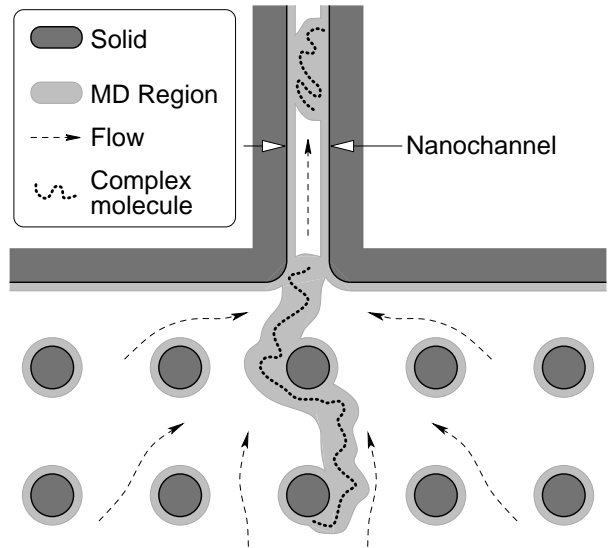}
    \caption{Schematic of an application of a hybrid MD/continuum simulation: complex molecules being transported into a nano channel.  Only the complex molecules, regions near them and regions near solid surfaces need an MD treatment, the remaining volume can be simulated with continuum mechanics.}
    \label{Figure_postsAndNanochannel}
\end{figure}

Some MD fluid dynamics simulations have been reported~\cite {
MD_Hirshfeld_Rapaport, MD_Rapaport_Clementi,
MD_Travis_Todd_Evans_Poiseuille, MD_Travis_Todd_Evans_Departure,
MD_Qian_Wang }, but MD is prohibitively computationally costly for
simulations of systems beyond a few tens of nanometers in size.
Fortunately, the molecular detail of the full flow-field that MD
simulations provide is often unnecessary; in liquids, beyond 5--10
molecular diameters ($\lesssim 3\mathrm{nm}$ for water) from a solid
surface the continuum-fluid approximation is valid and the
Navier-Stokes equations with bulk fluid properties may be used
\cite{NL_Koplik_Banavar,NL_Becker_Mugele,MD_Okumura_Heyes}.  Hybrid
simulations have been proposed~\cite {
    MD_Delgado-Buscalioni_Coveney_Phil_Trans_Royal_Soc,
    MD_Wagner_Flekkoy,
    MD_Nie_Chen_Robbins,
    MD_Werder_Walther_Koumoutsakos
} to simultaneously take advantage of the accuracy and detail
provided by MD in the regions that require it, and the computational
speed of continuum mechanics in the regions where it is applicable.
An example application of this technique is shown schematically in
figure~\ref{Figure_postsAndNanochannel}, where a complex molecule is
being electrokinetically transported into a nanochannel for
separation and identification\cite{NL_Pennathur_Santiago_2}.  Only
the complex molecule, its immediately surrounding solvent molecules,
and selected near-wall regions require an MD treatment; the
remainder of the fluid (comprising the vast majority of the volume)
may be simulated by continuum mechanics.  A hybrid simulation would
allow the effect of different complex molecules, solvent electrolyte
composition, channel geometry, surface coatings and electric field
strengths to be analysed, at a realistic computational cost.

In order to produce a useful, general simulation tool for hybrid
simulations, the MD component must be able to model complex
geometrical domains.  This capability does not exist in currently
available MD codes: domains are simple shapes, usually with periodic
boundaries.  The most important, computationally demanding, and
difficult aspect of any MD simulation is the calculation of
intermolecular forces.  This paper describes an algorithm that is
capable of calculating pair force interactions in arbitrary,
unstructured mesh geometries that have been parallelised for
distributed computing by spatial decomposition.

\subsection{Neighbour Lists are Unsuitable}

The conventional method of MD force evaluation in distributed
parallel computation is to use the cells algorithm to build
`neighbour lists' for interacting pairs with the `replicated
molecule' method of providing interactions across periodic
boundaries and interprocessor boundaries, where the system has been
spatially
partitioned~\cite{MD_Rapaport_art,MD_Smith_hypercube,MD_Rapaport_Million_II}.

The spatial location of molecules in MD is dynamic, and hence not
deducible from the data structure that contains them.  A neighbour
list defines which pairs of molecules are within a certain distance
of each other, and as such need to interact via intermolecular
forces.

\begin{figure}[htb]
    \centering
    \includegraphics[width=11cm]{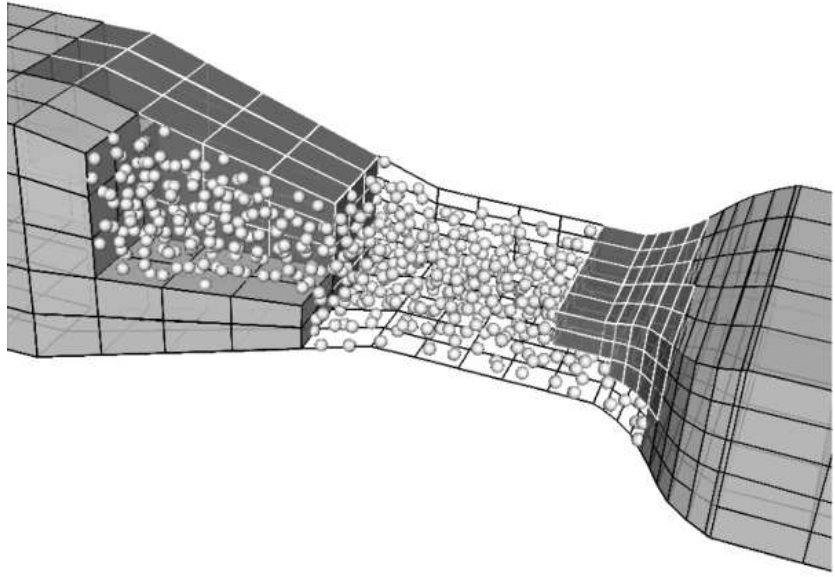}
    \caption{Example of a molecular dynamics flow simulation in a complex nano channel.  The mesh is decomposed into five irregular portions, four of which are shown as shaded blocks, the molecules contained in the fifth portion are shown.}
    \label{Figure_testMolconfig_complex_BW}
\end{figure}

\emph{Neighbour lists are unsuitable} when considering systems of
arbitrary geometry, that may have been divided into irregular and
complex mesh segments using standard mesh partitioning techniques
(see, for example, figure~\ref{Figure_testMolconfig_complex_BW}) for
two reasons:

\begin{enumerate}

\item \textbf{Interprocessor molecule transfers:} A molecule may cross an interprocessor boundary at any point in time (even part of the way through a timestep), at which point it should be deleted from the processor it was on and an equivalent molecule created on the processor on the other side of the boundary. Given that neighbour lists are constructed as lists of array indices, references or pointers to the molecule's location in a data structure, deleting a molecule would invalidate this location and require searching to remove all mentions of it.  Likewise, creating a molecule would require the appropriate new pair interactions to be identified.  Clearly this is not practical. It is conventional to allow molecules to `stray' outside of the domain controlled by a processor and carry out interprocessor transfers (deletions and creations) during the next neighbour list rebuild.  This is only possible when the spatial region associated with a processor can be simply defined by a function relating a position in space to a particular cell on a particular processor (i.e. a uniform, structured mesh, representing a simple domain).  In a geometry where the space in question is defined by a collection of individual cells of arbitrary shape, this is not possible.  For example the location the molecule has `strayed' to may be on the other side of a solid wall on the neighbour processor, or across another interprocessor boundary.  For the molecule to end up unambiguously in the correct place, interprocessor transfers must happen as molecules cross a face.

\item \textbf{Spatially resolved flow properties:} MD simulations used for flow studies must be able to spatially resolve fluid mechanic and thermodynamic fields.  This is achieved by accumulating and averaging measurements of the properties of molecules in individual cells of the same mesh that defines the geometry.  If a molecule is allowed to stray outside of the domain controlled by a processor, as above, then it would not be unambiguous and automatic which cell's measurement the molecule should contribute to.

\end{enumerate}

While both of these problems could be mitigated by, for example,
working out which cell a molecule outside the domain should be in on
another processor and sending its information, doing so would result
in an inelegant and inflexible arrangement with each additional
simulation feature (i.e.\ a different measured variable or class of
intermolecular potential) requiring special treatment.

We have, therefore, developed a new algorithm which is of comparable
computational cost to neighbour lists, but designed to be powerful
and generic for simulation in arbitrary geometries.  As will be
shown, neighbour lists also have some unfavourable features that
could be improved upon.

\section{Arbitrary Interacting Cells Algorithm (AICA)}

\label{Text_arbitraryInteractingCellsAlgorithm}

\subsection{Replicated Molecule Periodicity and Parallelisation}

\label{Text_replicatedMoleculePeriodicityAndParallelisation}

\begin{figure}[htb]
    \centering
    \includegraphics[width=10cm]{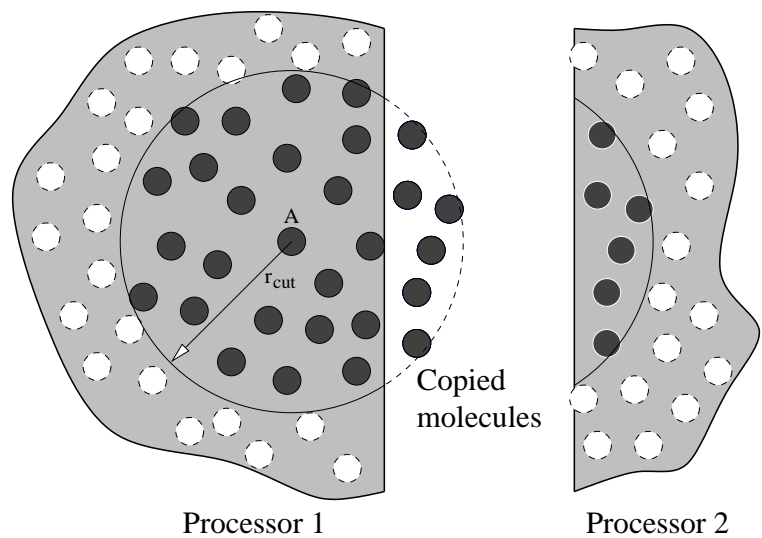}
    \caption{Domain decomposition. Molecule A must calculate intermolecular force contributions from all other molecules within its cut-off radius, $r_{cut}$.  When the domain has been decomposed, some of these molecules may lie on a different processor.  In this case, copies of the appropriate molecules from processor 2 are made on processor 1.}
    \label{Figure_referredMolecules}
\end{figure}

\begin{figure}[htb]
    \centering
    \includegraphics[width=7.2cm]{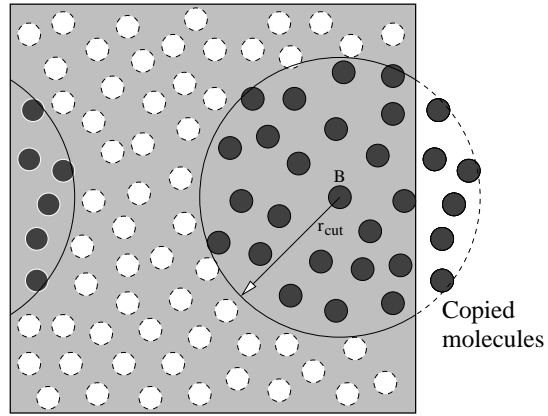}
    \caption{Periodic boundaries.  Molecule B must calculate intermolecular force contributions from all other molecules within $r_{cut}$. Some of those molecules may be on a periodic image of the system, and as such, in implementation terms, will reside on the other side of the domain.  Serial calculations in simple geometries typically use the minimum image convention~\cite{MD_Rapaport_art,MD_Allen_Tildesley}, but this is not suitable for parallelisation.}
    \label{Figure_referredMoleculesPeriodic}
\end{figure}

When parallelising an MD calculation, the spatial domain of the
simulation is decomposed and each processor is given responsibility
for a single region~\cite{MD_Rapaport_art}.  Molecules are, of
course, able to cross the boundaries between these regions and
therefore need to be communicated from one processor to the next.
Processors also communicate when carrying out intermolecular force
calculations, in which molecules close to processor boundaries need
to be replicated on their neighbours to provide interactions.  This
process is illustrated in figure~\ref{Figure_referredMolecules}.
Periodic boundaries also require information about molecules that
are not adjacent physically in the domain (see
figure~\ref{Figure_referredMoleculesPeriodic}), these required
interactions can also be constructed by creating copies of molecules
outside the boundary.

It is possible to handle processor and periodic boundaries in
exactly the same way, since they have the same underlying objective:
molecules near to the edge of a region need to be copied either
between processors or to other locations on the same processor at
every timestep to provide interactions.  This is a useful feature
because decomposing a mesh for parallelisation will often turn a
periodic boundary into a processor boundary.  The issue is how to
efficiently identify which molecules need to be copied, and to which
location, because this set continually changes as the molecules
move.

\subsection{Interacting Cell Identification}

\label{Text_interactingCellIdentification}

\begin{figure}[htb]
    \centering
    \includegraphics[width=7cm]{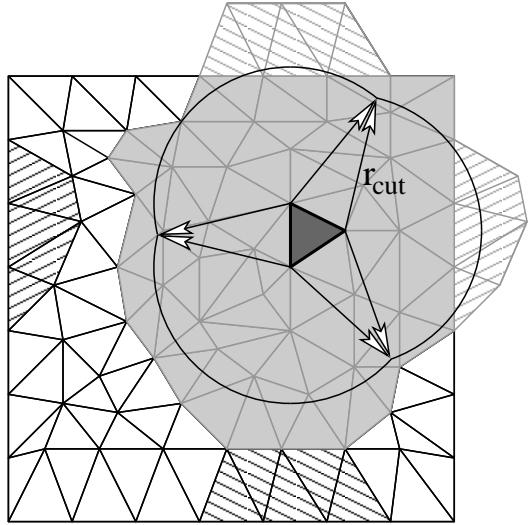}
    \caption{Interacting cell identification using unstructured triangular cells to demonstrate insensitivity to mesh topology.  The spatial domain (main square section comprising the real cells) is periodic top-bottom and left-right.  Real cells within $r_{cut}$ that interact with the CIQ (dark) are shaded in grey.  The required referred cells are hatched in alternate directions according to which boundary they have been referred across.  Realistic systems would be significantly larger compared to $r_{cut}$ than shown here.}
    \label{Figure_interactingCellIdentification}
\end{figure}

The new Arbitrary Interacting Cells Algorithm (AICA) we propose here
is an extension of the Conventional Cells Algorithm (CCA)
\cite{MD_Rapaport_art}.  In the CCA, a simple (usually cuboid)
simulation domain is subdivided into equally sized cells.  For
computational and theoretical reasons, intermolecular potentials do
not extend to infinity; they are assigned a cut-off radius,
$r_{cut}$, beyond which they are set to zero.  The minimum dimension
of the CCA cells must be greater than $r_{cut}$, so that all
molecules in a particular cell interact with all other molecules in
their own cell and with those in their nearest neighbour cells
(i.e.\ those they share a face, edge or vertex with --- 26 in 3D).
AICA uses the same type of mesh as would be used in Computational
Fluid Dynamics (CFD) to define the geometry of a region. \emph{There
are no restrictions on cell size, shape or connectivity.} Each cell
has a unique list of other cells that it is to interact with, this
list is known as the Direct Interaction List (DIL) for the Cell In
Question (CIQ).  It is constructed by searching the mesh to create a
set of cells that have at least one vertex within a distance of
$r_{cut}$ from any of the vertices of the CIQ, see
figure~\ref{Figure_interactingCellIdentification}.

The DILs are established prior to the start of simulation and are
valid throughout because the spatial relationship of the cells is
fixed, whereas the set of molecules they contain is dynamic. In a
similar way to the CCA, at every timestep a molecule in a particular
cell calculates its interactions with the other molecules in that
cell and consults the cell's DIL to find which other cells contain
molecules it should interact with.  Information is required to be
maintained stating which cell a molecule is in --- this is
straightforward and computationally cheap.

The construction of the DILs and accessing molecules is done in such
a way as to not double-count interactions, similarly to the CCA. For
example, if cells A and B need to interact, cell B will be on cell
A's DIL, but cell A will not be on cell B's DIL.  When a molecule in
cell A interacts with one in cell B the molecule in cell B will
receive the inverse of the force vector calculated to be added to
the molecule in cell A, because the pair forces are reciprocal,
i.e.\ $\mathbf{f_{ab}} = \mathbf{-f_{ba}}$.

\begin{figure}[htb]
    \centering
    \includegraphics[width=7cm]{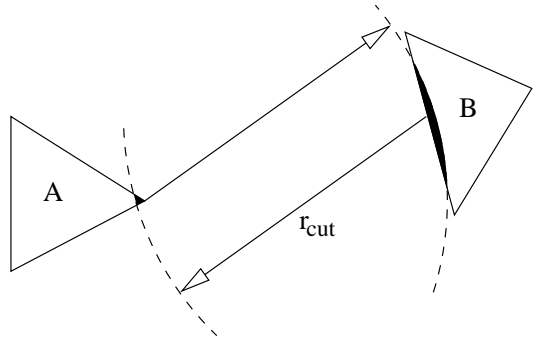}
    \caption{Errors caused by cut-off radius searching from vertices only.  An arc of radius $r_{cut}$ drawn from the indicated vertex on cell A intersects a face of cell B, therefore, molecules near this vertex should interact with molecules in the shaded region.  Similarly, an arc of radius $r_{cut}$ drawn from the indicated face on cell B encompasses the small shaded region of cell A. The cell searching algorithm will not, however, identify cells A and B as needing to interact because none of their vertices lie within $r_{cut}$ of each other.}
    \label{Figure_errorsInTwistedMeshes}
\end{figure}

It is possible in rare cases for small errors to be caused by this
algorithm: small slices of volume in cells may not be identified for
interaction (see figure~\ref{Figure_errorsInTwistedMeshes}). The
errors introduced will be slight because the relative volume not
accounted for will be very small: in
figure~\ref{Figure_errorsInTwistedMeshes}, only when molecules are
in both the indicated regions of each cell will either miss out on
an interaction, and the intermolecular potential will be small at
this distance. A small guard distance could be added to $r_{cut}$
when constructing the DIL to reduce this error. This is less of a
problem in good quality meshes, e.g.\ hexahedral rather than
tetrahedral cells.

\subsubsection{Referred Molecules and Cells}

\label{Text_referredMoleculesAndCells}

Replicated molecule parallelisation and periodic boundaries are
handled in the same way using \emph{referred cells} (see
figure~\ref{Figure_interactingCellIdentification}) and
\emph{referred molecules}.

\begin{description}

\item [Referred molecule: ]  A copy of a real molecule that has been placed in a region outside a periodic or processor boundary in order to provide the correct intermolecular interaction with molecules inside the domain.  A referred molecule holds only its own position and id (i.e.\ identification of which type of molecule it is for heterogenous simulations).  Referred molecules are created and discarded at each timestep, and do not report any information back to their source molecules.  Therefore if molecule \emph{j} on processor 1 needs to interact with molecule \emph{k} on processor 2, a separate referred molecule will be created on each processor.

\item [Referred cell: ]  Referred cells define a region of space and hold a collection of referred molecules.  Each referred cell knows
\begin{itemize}

\item which real cell in the mesh (on which processor) is its source;

\item the required transformation to refer the positions and orientations of the real molecules in the source cell to the referred location (see below for the details of how cells and molecules are referred);

\item the positions of all of its own vertices.  These are the positions of the vertices of the source cell which have been transformed by the same referring transform as the referred molecules it contains;

\item which real cells are in range of this particular referred cell and hence require intermolecular interactions to be calculated.  This is constructed once at the start of the simulation in the same way as the DIL for real-real cell interactions --- the vertices of the real cells in the portion of mesh on the same processor as the referred cell are searched, those with at least one vertex in range of any vertex of the referred cell are found.

\end{itemize}

\end{description}

\subsubsection{Referring Transformations}

\label{Text_cellPositionReferringTransformations}

\begin{figure}[htb]
    \begin{center}
    \input{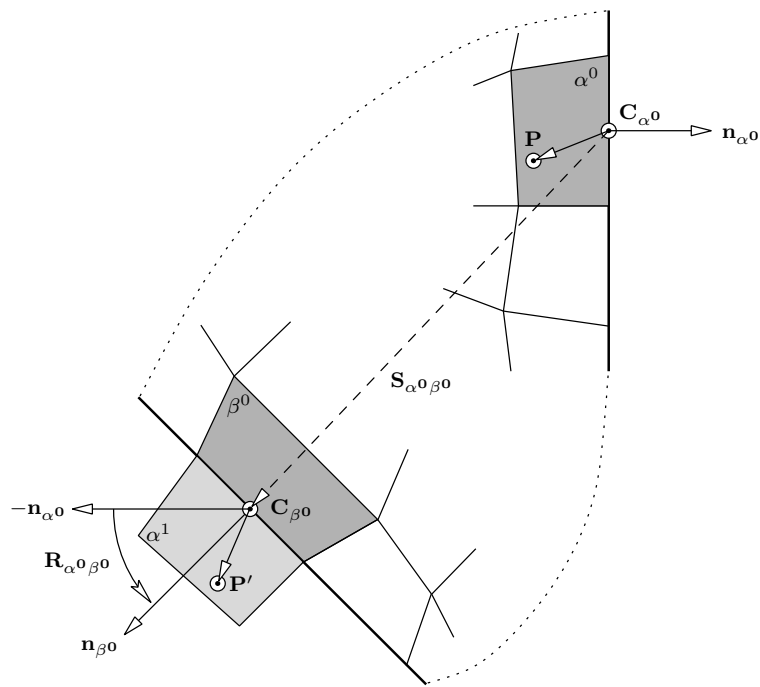}
    \caption{A molecule at point $\mathbf{P}$ is referred across the boundary (heavy line) to $\mathbf{P'}$ by a transformation defined by the face center/face normal vectors of the faces of cells $\alpha^0$ and $\beta^0$ on the boundary. The mesh is rigid, so all points of cell $\alpha^1$ have the same relative spatial relationship as $\alpha^0$.  The vertices of referred cell $\alpha^1$ are calculated by the same process.}
    \label{Figure_cellReferringTransform}
    \end{center}
\end{figure}

A spatial transformation is required to refer a cell across a
periodic or processor boundary.
Figure~\ref{Figure_cellReferringTransform} shows the most general
case of a cell being referred across a separated, non-parallel
boundary, where,

\begin{tabular}{r @{ = } l}

$\alpha^0$ & Cell with a face on one side of the boundary.\\

$\beta^0$ & Cell with a face on the other side of the boundary, coupled to $\alpha^0$\\

$\alpha^1$ & Cell $\alpha^0$ referred across the boundary.\\

$\mathbf{C}$ & Cell centre.\\

$\mathbf{n}$ & Face normal unit vector.\\

$\mathbf{S_{\alpha^0 \beta^0}}$ & $\mathbf{C_{\beta^0} - C_{\alpha^0}}$.  Position shift from $\mathbf{C_{\alpha^0}}$ to $\mathbf{C_{\beta^0}}$.\\

$\mathbf{R_{\alpha^0 \beta^0}}$ & Tensor required to rotate
$\mathbf{-\hat{n}_{\alpha^0}}$ to $\mathbf{\hat{n}_{\beta^0}}$,
given by,

\end{tabular}

\begin{equation}
\mathbf{
    R_{\alpha^0 \beta^0} =
    -
    \left(
        n_{\alpha} \cdot n_{\beta}
    \right) I}
    +
    \left(
        1 + \mathbf{n_{\alpha} \cdot n_{\beta}}
    \right)
    \mathbf{
    \left(
        \frac
{ n_{\alpha} \times n_{\beta} } { \left | n_{\alpha} \times
n_{\beta} \right | }
    \right)^2
    +
        n_{\alpha}n_{\beta}
    -
        n_{\beta}n_{\alpha}
}, \label{Equation_R_alpha0_beta1}
\end{equation}

\noindent where $\mathbf{I}$ is the identity tensor. The absolute
position, $\mathbf{P'}$, of a molecule has been transformed across a
boundary (with its containing cell) from its original position,
$\mathbf{P}$ is required. To derive the $\mathbf{P \rightarrow P'}$
transform, first the position of $\mathbf{P}$ relative to centre of
the coupled face on $\alpha^0$ is given as,

\begin{equation}
\mathbf{P - C_{\alpha^0}}.
\end{equation}

\noindent Then this vector is rotated by the transformation tensor
defined by the source and destination coupled face normals,

\begin{equation}
\mathbf{R_{\alpha^0 \beta^0} \cdot \left(P - C_{\alpha^0}\right)}.
\end{equation}

\noindent The rotated position is transformed back to global
coordinates,

\begin{equation}
\mathbf{C_{\alpha^0} + R_{\alpha^0 \beta^0} \cdot \left(P -
C_{\alpha^0}\right)},
\end{equation}

\noindent and then shifted by the relative separation of the coupled
face centres, i.e.\

\begin{align}
\mathbf{P'} & = \mathbf{
    C_{\alpha^0}
    +
    S_{\alpha^0 \beta^0}
    +
    R_{\alpha^0 \beta^0}
    \cdot
    \left(
        P - C_{\alpha^0}
    \right)
},
\nonumber \\
& = \mathbf{
    C_{\alpha^0}
    +
    C_{\beta^0}
    -
    C_{\alpha^0}
    +
    R_{\alpha^0 \beta^0}
    \cdot
    \left(
        P - C_{\alpha^0}
    \right)
},
\nonumber \\
& = \mathbf{
    C_{\beta^0}
    -
    R_{\alpha^0 \beta^0}
    \cdot
    C_{\alpha^0}
    +
    R_{\alpha^0 \beta^0}
    \cdot
    P
}. \label{Equation_referringResult}
\end{align}

The result is the same if the point is shifted by
$\mathbf{S_{\alpha^0 \beta^0}}$ prior to rotation by
$\mathbf{R_{\alpha^0 \beta^0}}$ around $\mathbf{C_{\beta^0}}$.  The
final position of  $\mathbf{P'}$ is the same if any coupled face
centre/face normal pair on the boundary is used.  Therefore, all
cells and all molecules referred across a particular boundary may
use the transformation derived from one cell.

This result can be used to transform the positions of all of the
molecules in a source cell to their correct position in a referred
cell. Molecules being referred also operate on their own
rotationally dependant properties (e.g.\ angular orientation) using
the rotation tensor. This transformation is suitable for multiple
periodic boundaries and arbitrary mesh decompositions where existing
referred cells must be re-referred by other boundaries.  This
creates the cell relationships across edges, corners, and also on
non-neighbouring processors. See
Appendix~\ref{Text_exampleConstructionOfReferredCells} for an
example.  Cell re-referring is achieved by writing
equation~\eqref{Equation_referringResult} as a generic transform,

\begin{equation}
\mathbf{P' = y + R \cdot P},
\end{equation}

\noindent where,

\begin{align}
\mathbf{y} &  = \mathbf{C_{\beta^0} - R_{\alpha^0 \beta^0} \cdot C_{\alpha^0}}, \\
\mathbf{R} &  = \mathbf{R_{\alpha^0 \beta^0}}.
\end{align}

\noindent If $\mathbf{y}$ and $\mathbf{R}$ are the transformations
required to move $\mathbf{P}$ to $\mathbf{P'}$, then transforming
$\mathbf{P'}$ to $\mathbf{P''}$ using $\mathbf{y'}$ and
$\mathbf{R'}$ gives,

\begin{align}
\mathbf{P''} & = \mathbf{y' + R' \cdot P'}, \nonumber \\
& = \mathbf{y' + R' \cdot y + R' \cdot R \cdot P}, \nonumber \\
& = \mathbf{y'^\star + R'^\star \cdot P},
\end{align}

\noindent reduced to the generic form by defining,

\begin{align}
\mathbf{y'^\star} & = \mathbf{y' + R' \cdot y},\\
\mathbf{R'^\star} & = \mathbf{R' \cdot R}.
\end{align}

\noindent Further re-referrals can be reduced to a single transform
in a predictable way, with only a single vector/tensor pair required
to be stored by the referred cell at any stage.  Any number of cell
referring transformations may be made sequentially, but the
resulting transformation takes the initial source position
($\mathbf{P}$) and directly transforms it to the final destination,
regardless of, and without having to intermediately visit, the other
referred locations along the way.

\subsection{Intermolecular Force Calculation Algorithm Details}

\label{Text_algorithmDetails}

At the start of the simulation, the DIL for each cell are created,
referred cells are created and they determine which real cells they
must supply interactions to.  Details of how these are constructed
can be found in Appendix~\ref{Text_cellInteractions}.  Cells
requiring to be referred need not be on processors sharing a
boundary with their destination processor, i.e.\ AICA must be able
to identify interactions between non-neighbouring processors.
However, when constructing the referred cells, processors may only
communicate across interprocessor faces, i.e.\ with neighbours only.

The geometry of the system is defined by a 3D, static, unstructured
mesh.  A cell's position in the data structure, hence its index,
does not imply anything about its physical location or connectivity.
Therefore, the ability to handle arbitrary geometries comes from
deducing locally which cells are in range to interact.  The same net
force on any molecule will be calculated irrespective of cell size
and shape (except in rare instances, see
section~\ref{Text_interactingCellIdentification}).

The total intermolecular force acting upon a molecule is calculated
at each timestep by considering two types of interactions.
\textbf{Real-Real} interactions occur between a molecule and others
on the same processor.  \textbf{Real-Referred} interactions occur
between real molecules and referred molecules arising from the other
side of a processor or periodic boundary.  Referred molecules do not
need to calculate interactions between themselves, because every
referred molecule is a copy of a real molecule elsewhere, and as
such will receive all of its real-real and real-referred
interactions in-situ.

\subsubsection{Real-Real Molecule Force Calculation}

At each time step, the real molecules on each processor calculate
their intermolecular interactions as follows:

\begin{verbatim}
for all cells, denoted I {
    for all molecules in cell I, denote mol k
    {
        for all cells in cell I's DIL, denote cell J
        {
            for all molecules in cell J, denote mol l
            {
                calculate intermolecular force, Fkl,
                between mol k and mol l.
                Add Fkl to mol k, add Flk = -Fkl to mol l.
            }
        }

        for all molecules in cell I, denote mol k'
        {
            if index of, or pointer to, mol k' > mol k
            then calculate intermolecular force, Fkk',
            between mol k and mol k'.
            Add Fkk' to mol k, add -Fkk' to mol k'.
        }
    }
}
\end{verbatim}

\noindent Comparison of the index, or pointer to molecules in the
same cell, \verb|mol k' > mol k|, means that interactions between
molecules in the same cell are not double counted and a molecule
does not calculate an interaction with itself.  The order of
comparison is not important, as long as all pairs are covered, so
using \verb|mol k' < mol k| as a condition would work equally well.

\subsubsection{Real-Referred Molecule Force Calculation}

At each timestep all real cells which are the source cell of one or
more referred cells send the position and id of all of the molecules
they contain to the appropriate referred cell on the appropriate
processor.  The destination referred cells perform the appropriate
position and orientation transformation when the molecules are
received. These referred molecules calculate their intermolecular
interactions with real molecules as follows:

\begin{verbatim}
for all referred cells, denoted M {
    for all referred molecules in referred cell M, denote mol q
    {
        for all real cells that referred cell M
        interacts with, denote cell N
        {
            for all real molecules in cell N, denote mol p
            {
                calculate intermolecular force, Fpq,
                between mol p and mol q.
                Add Fpq to mol p.
            }
        }
    }
}
\end{verbatim}

\subsection{Prediction of Computational Speed}

\label{Text_predictionsOfComputationalSpeed}

The following analysis determines whether or not AICA is practical,
by predicting its computational cost relative to that of the CCA and
Neighbour List Algorithm (NLA).

Assuming a system of approximately uniform density, the volume of
space around a particular molecule to be interrogated by a
particular algorithm will be directly proportional to the number of
pair interactions to be evaluated, which is itself directly
proportional the computational cost per timestep.  For a given cut
off radius, $r_{cut}$, the volume associated with a particular
algorithm is:

\begin{description}

\item[Ideal: ] Evaluating the minimum possible number of pairs is equivalent to a sphere of radius $r_{cut}$,

\begin{equation}
V_{ideal} = \frac{4}{3} \pi r_{cut}^3.
\end{equation}

\noindent However, it is not possible to design an algorithm that
only evaluates the ideal case because the relative spatial
arrangement of molecules is not known in advance from one timestep
to the next.

\item[CCA: ]  The volume encompassed by the CCA depends on the cell size --- it is minimum when cells are cubic with side length $r_{cut}$.  The volume of the cell containing the molecule in question, plus that of its 26 neighbours is encompassed, giving

\begin{equation}
V_{CCA} = 27 r_{cut}^3.
\end{equation}

\item[NLA: ] The neighbour list for a particular molecule includes all molecule interaction pairs within a sphere of radius $r_{cut}$ plus a guard distance $\Delta R$~\cite{MD_Rapaport_art}.  It is constructed using the CCA, where the cells must be at least $r_{cut} + \Delta R$ in size, and has a lifetime of $L$ timesteps before invalidation.  Therefore, each timestep that the neighbour list is used for must absorb a fraction of the CCA cost required to construct the list --- the shorter the lifetime of the neighbour list, the more costly the NLA becomes.  Therefore,

\begin{equation}
V_{NLA} = \frac{4}{3} \pi \left(r_{cut} + \Delta R \right)^3 +
\frac{27}{L} \left(r_{cut} + \Delta R \right)^3.
\end{equation}

\item[AICA: ] The volume encompassed by AICA is the total volume of all cells which have at least one vertex within $r_{cut}$ of any vertex of the cell containing the molecule in question, plus the volume of this cell.  This depends on the local topology of the mesh.  As a simple example, consider a uniform mesh of cubes of side length $w$.  A piecewise function, $N_c(r_{cut}/w)$ has been derived to count the number of cubes, each of volume $w^3$, that have at least one corner within a distance of $r_{cut}/w$ of a corner of a central cube (which is included).  In this case,

\begin{equation}
V_{AICA} = N_c(r_{cut}/w) w^3.
\end{equation}

\end{description}

\begin{figure}[htb]
    \centering
    \includegraphics[angle=-90,width=12cm]{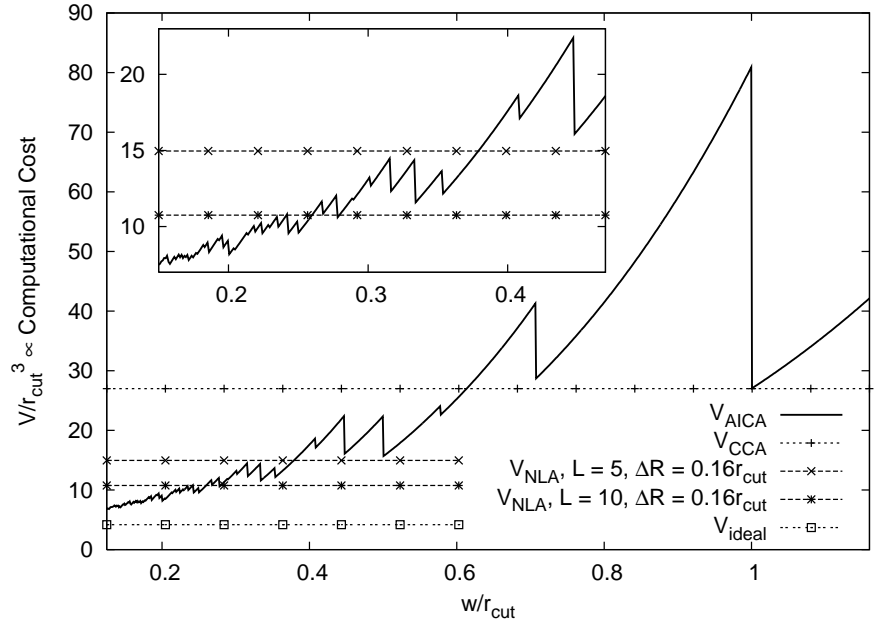}
    \caption{Predicted computational cost of new AICA using cubic cells, compared with conventional algorithms.  $\Delta R = 0.16r_{cut}$ derived from typical values for liquid density simulation $r_{cut} = 2.5$, $\Delta R = 0.4$~\cite{MD_Rapaport_art}.}
    \label{Figure_V_dimless_jComputPhys}
\end{figure}

\noindent These volumes are plotted on
figure~\ref{Figure_V_dimless_jComputPhys} with $w/r_{cut}$ as the
independent variable, lower values of $w/r_{cut}$ representing a
finer mesh for a given cut-off radius. AICA is costlier than the CCA
or NLA in coarse meshes, however, when $w \lesssim 0.4r_{cut}$ (see
inset for detail) AICA has a similar computational cost to the NLA,
for the chosen representative values of $L$ and $\Delta R$. The
discontinuities in $V_{AICA}$ are due to the steps in
$N_c(r_{cut}/w)$ which arise when a threshold is crossed and another
`layer' of cubes is required.

While real meshes in complex geometries will not comprise uniform
cubes, this result demonstrates that AICA can be of comparable speed
to the NLA, and gives an indication of the required refinement.  As
the mesh becomes finer, the set of cells selected to interact with
the cell at the centre becomes an increasingly accurate
approximation of the ideal sphere, this can be seen in the
$V_{AICA}$ function, which asymptotically approaches $V_{ideal}$ as
$w/r_{cut} \rightarrow 0$.  Note that, in this example, AICA and the
CCA are equivalent when $w/c_{cut} = 1$ because their cells are the
same size.

\subsubsection{Neighbour List Lifetime Prediction}

\label{Text_neighbourListLifetimePrediction}

The computational cost of the neighbour list algorithm depends on
the number of timesteps a neighbour list remains valid for --- a
quantity whose dependence on simulation properties can be predicted.

For a given number of molecules, $N$, of mass $m$, in equilibrium at
a temperature $T$, there is a probability of $1/N$ that a molecule
in the system will be travelling with a velocity $v_N$.  It is
therefore likely that at every timestep there will be \emph{one}
molecule travelling at, or close to this velocity.  Given that a
neighbour list is invalidated when the cumulative sum of
\emph{maximum} displacements in the system exceeds a fixed threshold
(usually $0.5 \Delta R$, half the guard
radius~\cite{MD_Rapaport_art}), finding the number of timesteps
required for a molecule travelling at $v_N$ to cover this distance
gives an estimate of the lifetime of the list.

\begin{figure}[htb]
    \centering
    \includegraphics[angle=-90,width=10cm]{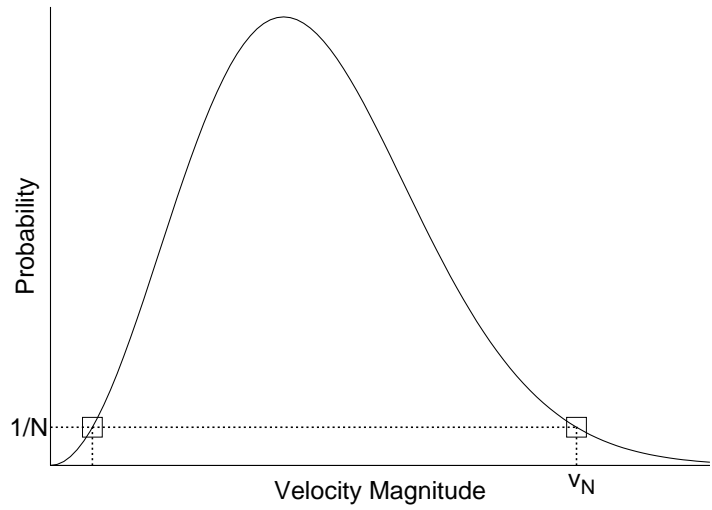}
    \caption{Maxwellian velocity distribution.  Finding velocities corresponding to a probability of $1/N$, where $N$ is the number of molecules in the system.  There are two real, positive solutions --- the high speed one is $v_N$, the quantity of interest.}
    \label{Figure_maxwellianVelocityDistribution}
\end{figure}

Finding $v_N$ by equating the Maxwellian velocity distribution to
$1/N$, where $T$, $m$ and $v_N$ are in reduced MD
units~\cite{MD_Allen_Tildesley} (see
figure~\ref{Figure_maxwellianVelocityDistribution}):

\begin{equation}
4\pi \left( \frac{m}{2 \pi T} \right)^{\frac{3}{2}} v_N^2
e^{\left(\frac{-m v_N^2}{2 T} \right)} = \frac{1}{N}.
\label{Equation_maxwellianVelocityDistibution_1_over_N}
\end{equation}

\noindent The high speed solution is,

\begin{equation}
v_N = \sqrt{-\frac{2 T}{m} W_{-1}\left(-\frac{1}{4 N} \sqrt{\frac{2
\pi T}{m}}\right)}, \label{Equation_vN_lambertW}
\end{equation}

\noindent where $W_{-1}$ is the secondary real branch of the Lambert
W function.  Therefore, $L$, the number of timesteps (of length
$\Delta t$) a neighbour list is valid for, is given by

\begin{equation}
L = \left\lceil\frac{0.5 \Delta R}{\Delta t v_N}\right\rceil.
\label{Equation_NLlifetime}
\end{equation}

\begin{figure}[htb]
    \centering
    \includegraphics[angle=-90,width=12cm]{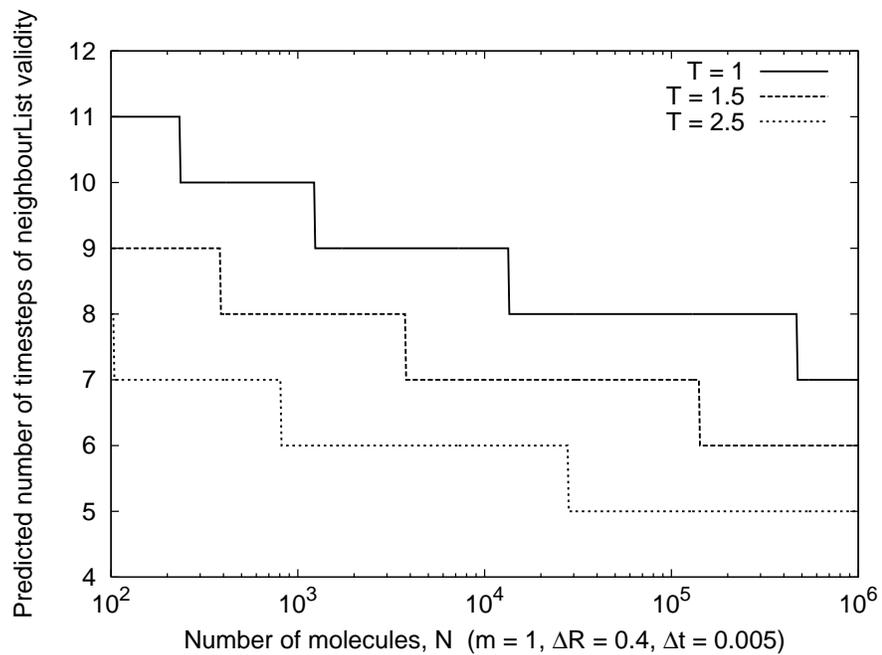}
    \caption{Variation of estimated neighbour list lifetime with typical simulation parameters for a Lennard Jones fluid~\cite{MD_Rapaport_art}. $T, m, \Delta R$ and $\Delta t$ are in reduced MD units.}
    \label{Figure_NL_life_steps_temperature_vary_quantised}
\end{figure}

\noindent This lifetime reduces with increasing temperature and
system size, as shown in
figure~\ref{Figure_NL_life_steps_temperature_vary_quantised},
resulting in a higher computational cost for the NLA.  At higher
temperatures, $v_N$ is higher because the velocity distribution
covers a higher molecular speed range.  In systems comprising more
molecules, it is more probable that at least one molecule will be
travelling at a given speed in the upper region of the distribution,
therefore $v_N$ is also higher.

The cost of AICA is not sensitive to either of these parameters,
therefore it becomes increasingly attractive in large systems
($>10^6$ molecules) where neighbour list lifetimes are low.

\section{Implementation and Verification}

\label{Text_implementationAndVerification}

AICA has been implemented in
OpenFOAM~\cite{MP_Weller_Tabor_Jasak_Fureby,MP_openFOAM}, which is
an open source C++ library intended for continuum mechanics
simulation of user-defined physics (primarily used for CFD) in
arbitrary, unstructured geometries.  The AICA MD code has been built
using OpenFOAM's lagrangian particle tracking library, which has
been used to track particles in solution, droplets in combusting
diesel sprays and to perform DSMC
simulations~\cite{MD_Allen_Hauser_foamDSMC}.

The AICA algorithm must show that it produces the correct results,
assessed by momentum and energy conservation and the behaviour of
average potential and kinetic energy, in three ways:

\begin{enumerate}

\item The code must produce the same results as another (validated) MD code.  For these purposes the other code is an in-house MD code which has been validated against those supplied in\cite{MD_Rapaport_art}.

\item The results for the AICA MD code must be the same whether it is run on one processor as a serial calculation, or run in parallel on many distributed CPUs.

\item The topology of the underlying mesh must not make any difference to the results.

\end{enumerate}

\begin{figure}[htb]
    \centering
    \includegraphics[width=10cm]{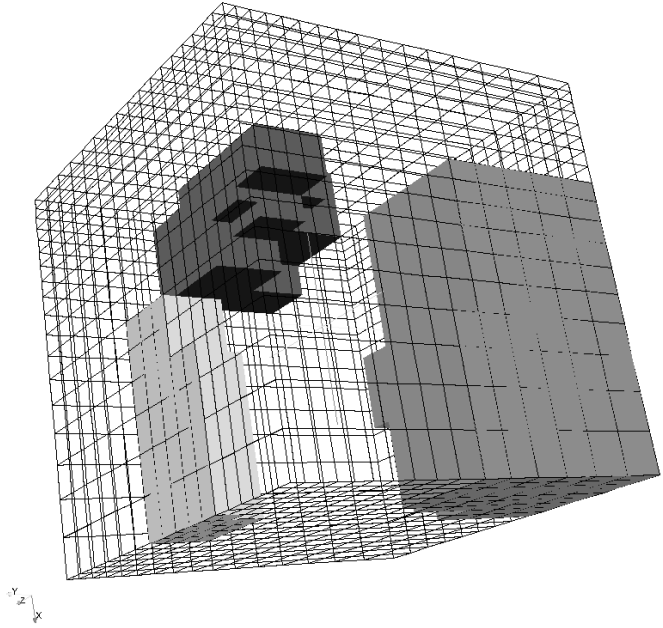}
    \caption{Mesh for the serial vs. parallel test; note the grading of cells.  The serial test uses the whole mesh (wireframe) and the parallel test decomposes the mesh into 24 portions (3 of which have been shown).}
    \label{Figure_testgnemdFOAM_gradedmesh_3_of_24_procs}
\end{figure}

The serial/parallel test case for AICA is shown in
figure~\ref{Figure_testgnemdFOAM_gradedmesh_3_of_24_procs}.  The
test mesh is decomposed into 24 portions, the three examples shown
are not regular shapes and the cells are nonuniform in size (the
mesh is graded in each direction).  While this test mesh is
structured, the algorithm makes no use of this fact, so a mesh
comprising cells of any size, shape or connectivity would produce
the same result.  This case will therefore demonstrate the
independence of the AICA from the form of the geometry used.

\begin{figure}[htb]
    \centering
    \includegraphics[angle=-90,width=12cm]{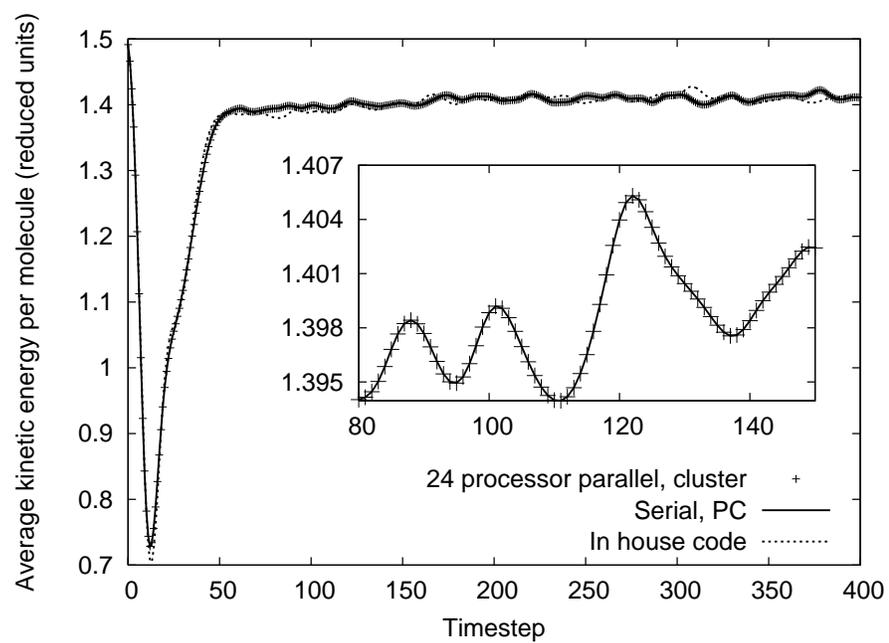}
    \caption{Verification of kinetic energy results.  Serial and parallel results are identical, see inset.}
    \label{Figure_verification_KE}
\end{figure}

\begin{figure}[htb]
    \centering
    \includegraphics[angle=-90,width=12cm]{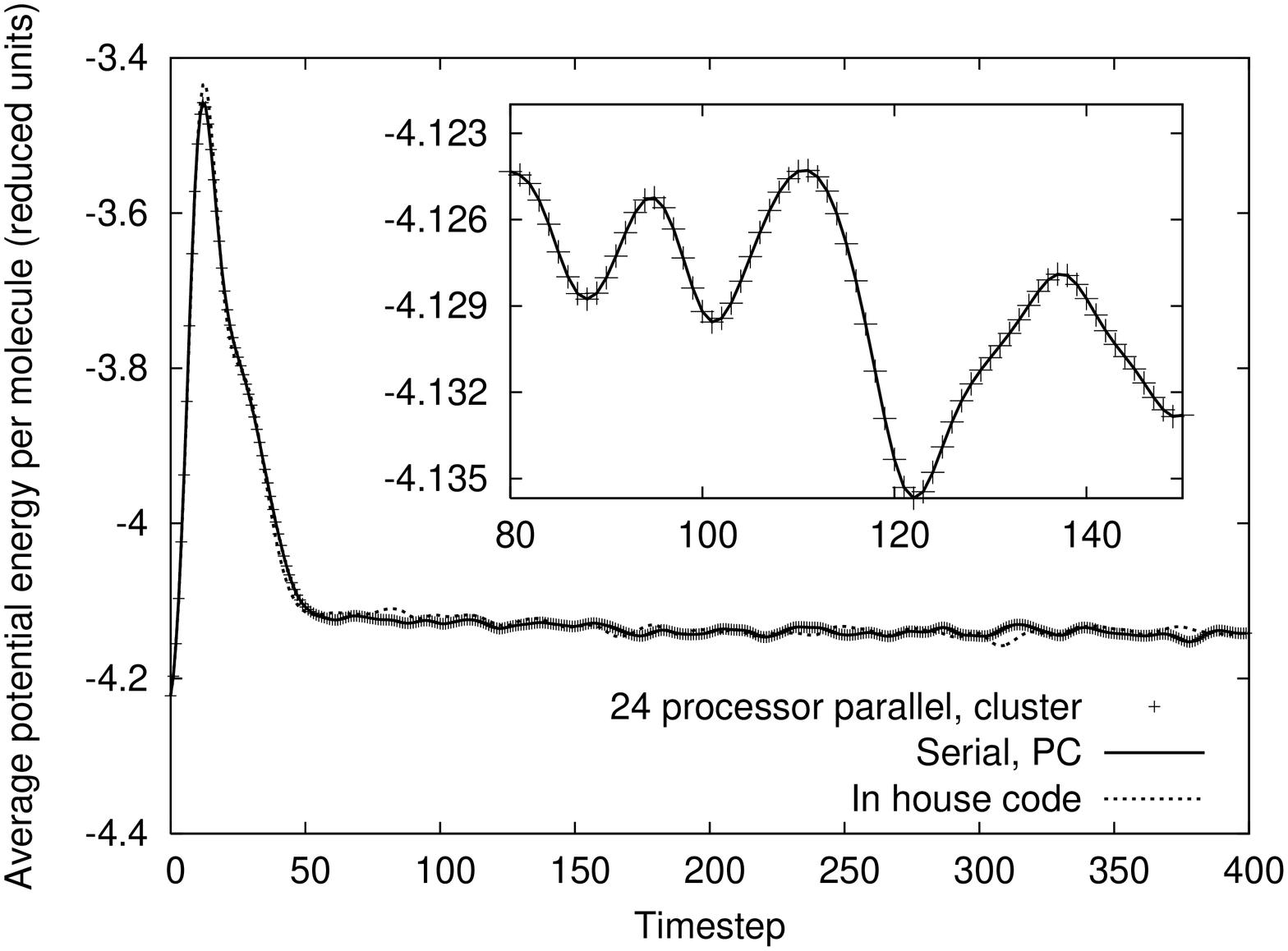}
    \caption{Verification of potential energy results.  Serial and parallel results are identical, see inset.}
    \label{Figure_verification_PE}
\end{figure}

\begin{figure}[htb]
    \centering
    \includegraphics[angle=-90,width=12cm]{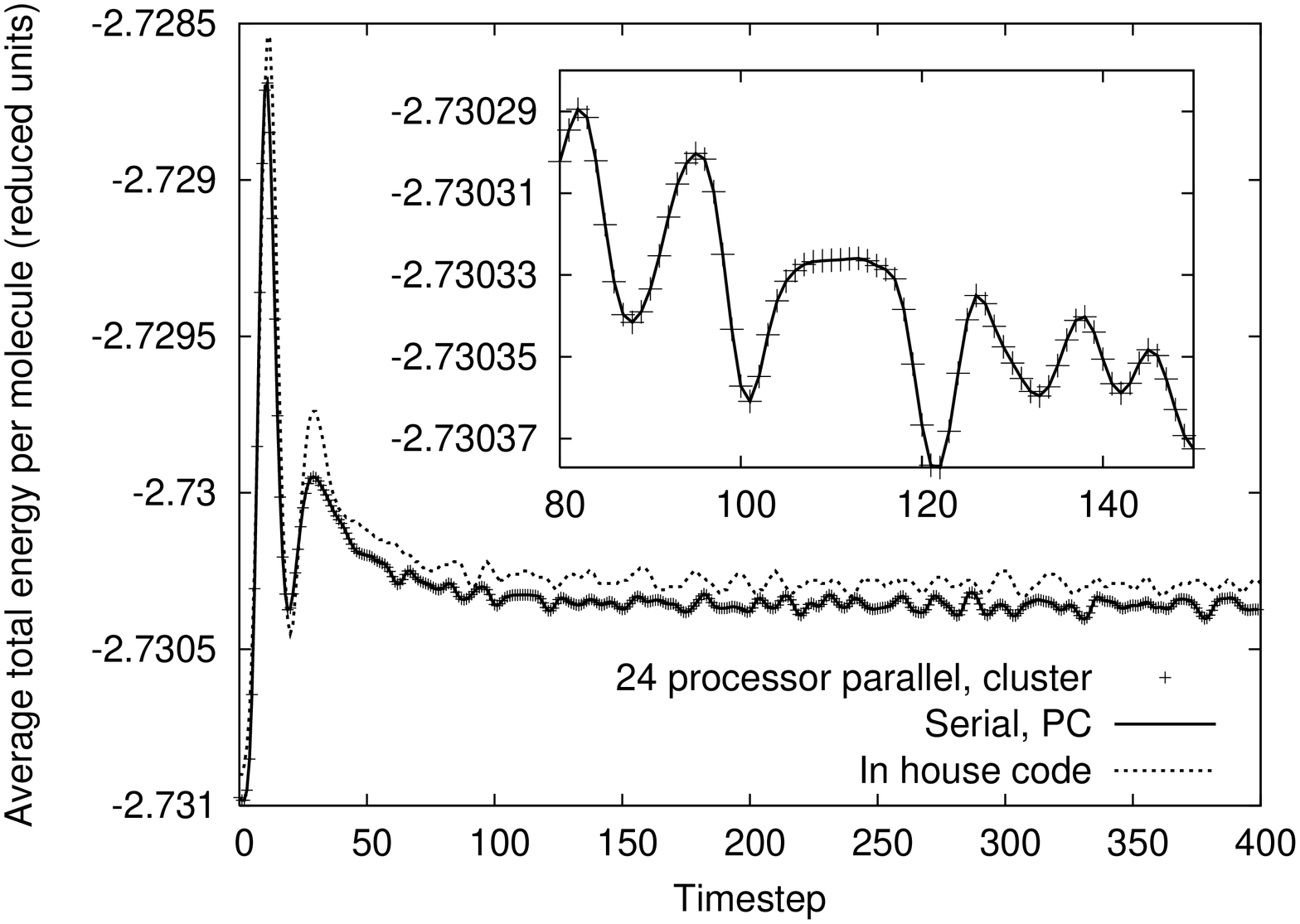}
    \caption{Verification of total energy results.  Serial and parallel results are identical, see inset.}
    \label{Figure_verification_TE}
\end{figure}

Figures~\ref{Figure_verification_KE}, \ref{Figure_verification_PE}
and \ref{Figure_verification_TE} show the average kinetic (KE),
potential (PE) and total (TE) energy per molecule for a system
containing 27000 Lennard Jones molecules, starting in a simple cubic
lattice at a temperature of 120K.  No equilibration or thermostat
has been applied.  Results for the in-house code, the AICA code in
serial on a PC and the AICA code decomposed and run on 24 processors
of a cluster show that the KE and PE graphs follow very similar
trends to the in-house code.  The simulations do not follow exactly
the same trajectory because the velocities in the systems are
initialised by random numbers, which are different between the two
codes.  This is likely to account for the slight (0.0025\%)
difference in average TE. The serial and parallel tests use the same
initial velocity configuration and each value for KE, PE and TE are
the same to better than 1 part in $10^{12}$.  Momentum is conserved
in both serial and parallel tests to approximately 1 part in
$10^{16}$.

Given that the serial and parallel results are \emph{exactly} the
same, the AICA must be creating the correct referred cells and
molecules, and calculating real-referred interactions perfectly,
even with the relatively complex decomposed portions of the mesh.

A comparison of how the computational speed changes with mesh
refinement, compared to the predictions in
section~\ref{Text_predictionsOfComputationalSpeed}, and further
tests of how the AICA performance scales in parallel will appear in
a future publication.

\section{Conclusions}

Our Arbitrary Interacting Cell Algorithm (AICA) has been described,
which is a new way of calculating pair forces in a molecular
dynamics simulation, carried out on an arbitrary unstructured mesh
that has been spatially decomposed to run in parallel.  The
algorithm is expected to achieve a similar computational speed to
the neighbour list method, but does not suffer from the identified
degradations in performance that neighbour lists experience in large
systems and at high temperatures.  AICA has been implemented in the
OpenFOAM code, and produces the same results when a system is
simulated in serial, or in parallel on a mesh decomposed into
irregular shapes on 24 processors.  The exchange of intermolecular
force information across interprocessor boundaries must therefore be
correct.

Currently this algorithm deals only with short ranged pair forces.
Future developments will include rigid and flexible polyatomic
molecules, and long-range electrostatic forces. A hybrid
MD/continuum implementation in OpenFOAM is currently under
development.  OpenFOAM is ideal for incorporating polar fluids and
electrokinetic actuation, which feature in envisioned applications,
because it is able to simulate electromagnetic, magnetohydrodynamic
and fluid mechanic continuum fields on the same mesh that AICA
operates with.

\section{Acknowledgements}

The authors would like to thank Chris Greenshields and Matthew Borg
of Strathclyde University (UK), and Henry Weller and Mattijs
Janssens of OpenCFD Ltd. (UK) for useful discussions. This work is
funded in the UK by Strathclyde University and the Miller
Foundation, and through a Philip Leverhulme Prize for JMR from the
Leverhulme Trust.

\newpage

\begin{appendix}

\section{Cell Interactions}

\label{Text_cellInteractions}

Building interaction lists and identifying and creating referred
cells is allowed to be computationally expensive (within reason)
because it will only happen once at the beginning of the MD
simulation.  Accessing the information, however, must be as fast as
possible because it will happen at every timestep.

Cell interactions are based on the largest cut-off radius of any
intermolecular potential in a heterogenous system, $\hat{r}_{cut}$.
It is possible to have different interaction lists when potentials
with substantially different cut-off radii are used.

\subsection{Building DILs}

\label{Text_realReferredCellInteractionsBuildDILs}

To create each real cell's DIL, a non double counting loop through
all cells is used, where the cells are accessed by an index ($i$ or
$j$) running $0 \rightarrow k-1$, where there are $k$ cells in the
portion of mesh on the processor in question, i.e.\

\begin{verbatim}
for cells, denote cell i, starting i = 0, while i < k-1 {
    for cells, denote cell j, starting j = i+1, while j < k
    {
        calculate magnitude of spatial separation of all
        vertices of cell i with all vertices of cell j.

        if any separation distance is less than the
        maximum cut off radius, add cell j to cell i's DIL.
    }
}
\end{verbatim}

\noindent For example, if there are $k = 5$ cells, the array index
runs from $0 \rightarrow 4$:

\begin{center}
\begin{tabular}{r | l}
$i$ & $j$ \\
\hline
$0$ & $1, 2, 3, 4$ \\
$1$ & $2, 3, 4$ \\
$2$ & $3, 4$ \\
$3$ & $4$ \\
\end{tabular}
\end{center}

\noindent All cell combinations have been created once, i.e.\ when
$i = 2$, $j = 1$ would produce the 2--1 combination, which is
already identified as 1--2. This non double counting procedure means
that some cells will have few or no other cells in their DIL,
however, as long as the correct cell pair is created in one of the
cell's DIL, this is not an issue.  It is also designed so that a
cell does not end up on its own DIL, the interactions between
molecules in the same cell are handled separately.

\subsection{Creating Referred Cells}

\label{Text_realReferredCellInteractionsCreateReferredCells}

\emph{Coupled patches} are the basis of periodic and interprocessor
communication for creating referred cells.  Patches, in general, are
a collection of cell faces representing a mesh boundary of some
description --- they may provide solid surfaces, inlets, outlets,
symmetry planes, periodic planes, or interprocessor connections.
Coupled patches provide two surfaces; whatever exits one enters the
other and vice versa. Two types are used in AICA:

\begin{itemize}

\item \emph{Periodic patches} on a single processor are arranged into two halves, each half representing one of the coupled periodic surfaces.  When a molecule crosses a face on one surface, it is wrapped around to the corresponding position on the corresponding face on the other surface.

\item \emph{Processor patches} provide links between portions of the mesh on different processors, one half of the patch is on each processor.  When a molecule crosses a face on one surface, it is moved to the corresponding position on the corresponding face on the other surface, on the other processor.  Decomposing a mesh for parallelisation will often require a periodic patch to be changed to a processor patch.

\end{itemize}

The surfaces of coupled patches may have any relative orientation,
and may be spatially separated as long as the face pairs on each
surface correspond to each other.  This allows them to also be used
as a `recycling' surface in a flow simulation --- whatever exits
from an outlet is reintroduced at an inlet, with the option of
scaling the molecule's properties, e.g.\ temperature and pressure.

\subsubsection{Create Patch Segments.}

In the decomposed portion of the mesh on each processor, each
processor and periodic patch should be split into segments, such
that:

\begin{itemize}

\item faces on a processor that were internal to the mesh prior to decomposition end up on a segment;

\item faces on a processor patch that were on separate periodic patches on the undecomposed mesh end up on different segments.  These segments are further split such that faces that were on different halves of the periodic patch on the undecomposed mesh end up on different segments;

\item faces on different halves of a periodic patch end up on different segments.

\end{itemize}

Each segment must produce a single vector/tensor transformation pair
(see section~\ref{Text_cellPositionReferringTransformations}) which
will be applied to all cells referred across it.

\subsubsection{Cell Referring Iterations.}

For each patch segment:

\begin{enumerate}

\item Find all real and existing referred cells in range (with at least one vertex within a distance of $\hat{r}_{cut}$) of at least one vertex of the faces comprising the patch segment.

\item Refer or re-refer this set of cells across the boundary defined by the patch segment using its transformation, see section~\ref{Text_cellPositionReferringTransformations}.  In the case of a segment of periodic boundary, this creates new referred cells on the same processor.  For a segment of a processor patch, these cells are communicated to, and created on the neighbouring processor.  Before creating any new referred cell a check is carried out to ensure that it is not

    \begin{itemize}

    \item a duplicate referred cell, one that has been created already by being referred across a different segment;

    \item a referred cell trying to be duplicated on top of a real cell, i.e. a cell being referred back on top of itself.

    \end{itemize}

\noindent If the proposed cell for referral would create a duplicate
of an existing referred cell, or end up on top of a real cell, then
it is not created.  To be a duplicate, the source processor, source
cell and the vector part of the transformation (see
section~\ref{Text_cellPositionReferringTransformations}) must be the
same for the two cells (note, the vector part of transformation for
a real cell is zero by definition).

\end{enumerate}

A single run of these steps will usually not produce all of the
required referred cells.  They are repeated until no processor adds
a referred cell in a complete evaluation of all segments, meaning
all possible interactions are accounted for.  In iterations after
the first, in step 1 it is enough to only search for referred cells
in range of the faces on the patch segment, because the real cells
will not have changed, and would all be duplicates.  The final
configuration of referred cells does not depend on the order of
patch segment evaluation.
Appendix~\ref{Text_exampleConstructionOfReferredCells} contains an
example of the results of the cell referring process.

\section{Example Construction of Referred Cells}

\label{Text_exampleConstructionOfReferredCells}

\begin{figure}[htb]
    \centering
    \includegraphics[width=8cm]{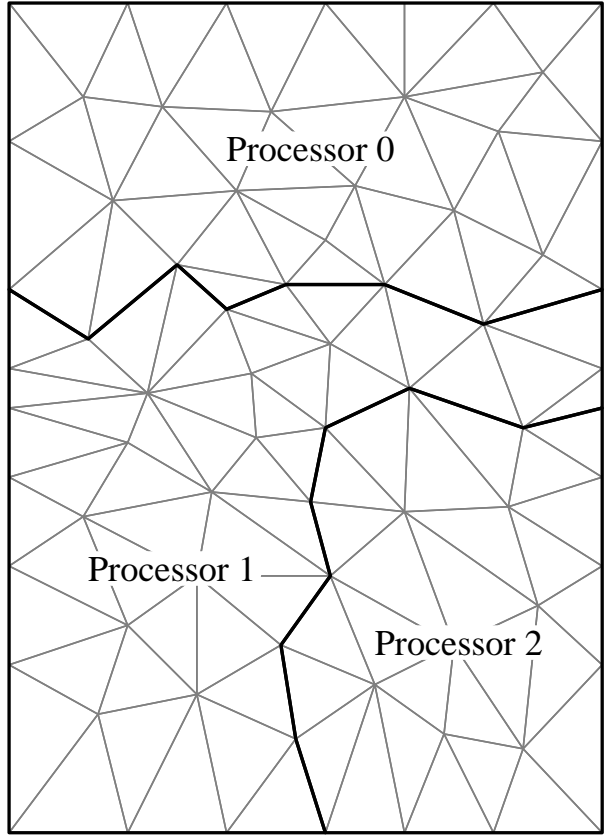}
    \caption{An arbitrary unstructured mesh, which is periodic top-bottom and left-right.  It is decomposed into 3 irregular portions for parallel processing as marked.}
    \label{Figure_parallelGridPolyhedra1Step_BW}
\end{figure}

\begin{figure}[htb]
    \centering
    \includegraphics[width=14cm]{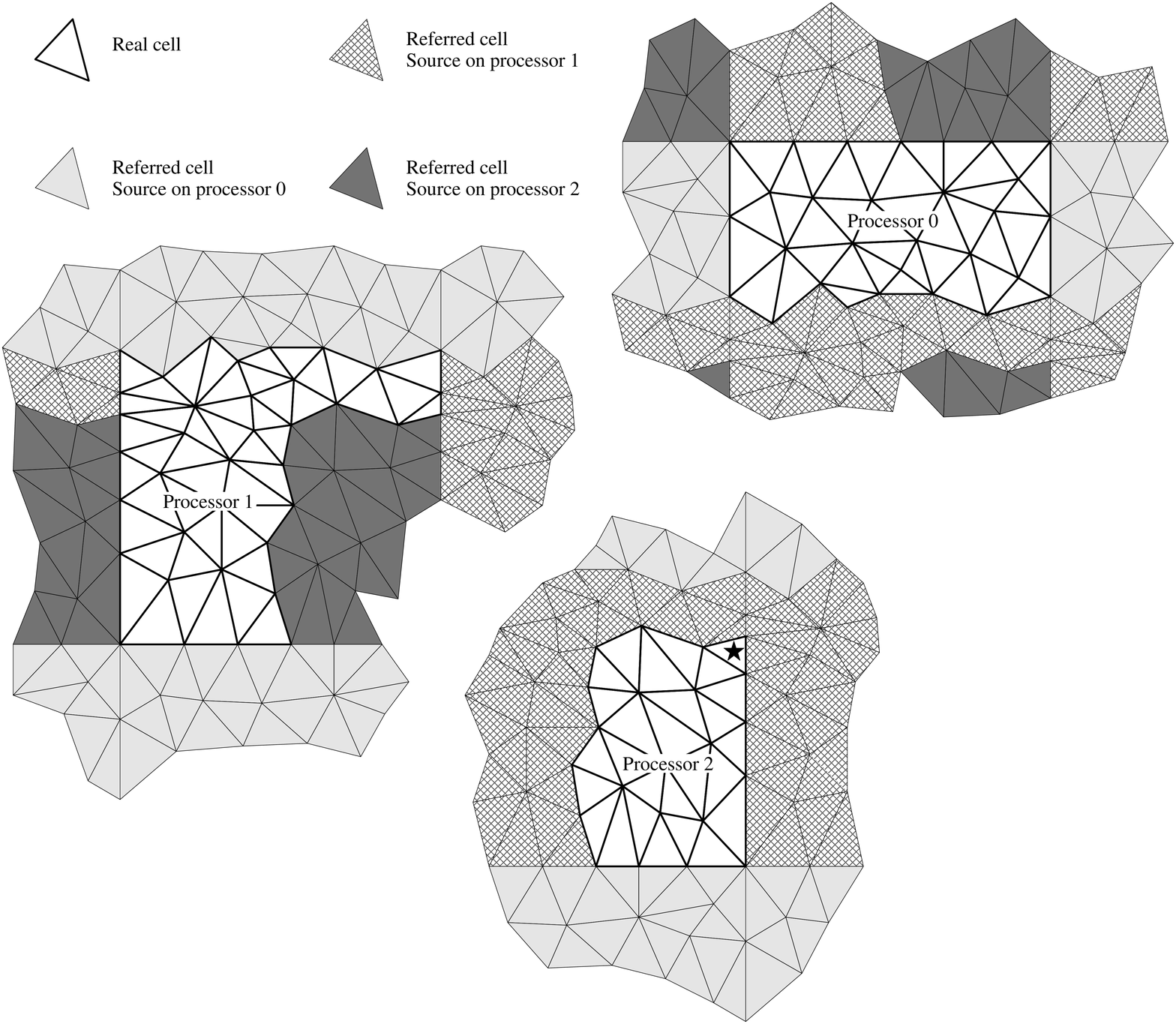}
    \caption{The final configuration of referred cells around the portions of the mesh in figure~\ref{Figure_parallelGridPolyhedra1Step_BW} on each processor.  A circle of radius $\hat{r}_{cut}$ drawn from any vertex of a real cell would be fully encompassed by other real or referred cells, thereby providing all molecules in that cell with the appropriate intermolecular interactions, either across a periodic boundary or from another processor.  Note that many cells are referred several times; for example, the source cell marked with $\bigstar$ on processor 2 is referred to processors 0 and 1 twice on each.}
    \label{Figure_parallelGridPolyhedra2Step_BW}
\end{figure}

Figures \ref{Figure_parallelGridPolyhedra1Step_BW} and
\ref{Figure_parallelGridPolyhedra2Step_BW} show the start and end
points of the cell referring operation on a mesh that has been
decomposed to run in parallel.  The example is in 2D for clarity but
the process is exactly the same in 3D.

In this example the number of referred cells created far exceeds the
number of real cells, which would lead to much costly interprocessor
communication.  This is because the mesh has been made deliberately
small relative to $\hat{r}_{cut}$ to demonstrate as many features of
the algorithm as possible and to be practical to understand.  In
realistic systems, the mesh portions would be significantly bigger
than $\hat{r}_{cut}$ and the referred cells would form a relatively
thin `skin' around each portion.  Decomposition of the mesh should
preferably be carried out to minimise the number of cells that need
to be referred, and to ensure that the vast majority of the
intermolecular interactions happen between real-real molecule pairs;
in this way the communication cost is minimised.

\end{appendix}

\newpage

\newpage

\newpage

\newpage

\newpage

\newpage

\newpage

\newpage

\newpage

\newpage

\newpage

\newpage

\newpage

\newpage

\newpage

\newpage

\newpage

\begin{thebibliography}{22}

\bibitem{NL_Agre_Aquaporin}
P.~Agre, {The aquaporin water channels}, Proceedings of the American Thoracic
  Society 3~(1) (2006) 5--13.

\bibitem{NL_Koplik_Banavar}
J.~Koplik, J.~R. Banavar, Continuum deductions from molecular hydrodynamics,
  Annual Review of Fluid Mechanics 27 (1995) 257--292.

\bibitem{MD_Brenner_Ganesan}
H.~Brenner, V.~Ganesan, Molecular wall effects: are conditions at a boundary
  ``boundary conditions''?, Physical Review E 61~(6) (2000) 6879--6897.

\bibitem{MD_Hirshfeld_Rapaport}
D.~Hirshfeld, D.~Rapaport, Molecular dynamics simulation of {T}aylor-{C}ouette
  vortex formation, Physical Review Letters 80~(24) (1998) 5337--5340.

\bibitem{MD_Rapaport_Clementi}
D.~Rapaport, E.~Clementi, Eddy formation in obstructed fluid flow: a
  molecular-dynamics study, Physical Review Letters 57~(6) (1986) 695--698.

\bibitem{MD_Travis_Todd_Evans_Poiseuille}
K.~Travis, B.~Todd, D.~Evans, Poiseuille flow of molecular fluids, Physica A
  240~(1-2) (1997) 315--27.

\bibitem{MD_Travis_Todd_Evans_Departure}
K.~Travis, B.~Todd, D.~Evans, Departure from {N}avier-{S}tokes hydrodynamics in
  confined liquids, Physical Review E 55~(4) (1997) 4288--95.

\bibitem{MD_Qian_Wang}
T.~Qian, X.-P. Wang, Driven cavity flow: from molecular dynamics to continuum
  hydrodynamics, Multiscale Modeling and Simulation 3~(4) (2005) 749--763.

\bibitem{NL_Becker_Mugele}
T.~Becker, F.~Mugele, Nanofluidics: viscous dissipation in layered liquid
  films, Physical Review Letters 91 (2003) 166104.

\bibitem{MD_Okumura_Heyes}
H.~Okumura, D.~M. Heyes, Comparisons between molecular dynamics and
  hydrodynamics treatment of nonstationary thermal processes in a liquid,
  Physical Review E 70~(6) (2004) 061206.

\bibitem{MD_Delgado-Buscalioni_Coveney_Phil_Trans_Royal_Soc}
R.~Delgado-Buscalioni, P.~V. Coveney, Hybrid molecular-continuum fluid
  dynamics, Philosophical Transactions of the Royal Society London 362~(1821)
  (2004) 1639--1654.

\bibitem{MD_Wagner_Flekkoy}
G.~Wagner, E.~G. Flekk{\o}y, Hybrid computations with flux exchange,
  Philosophical Transactions of the Royal Society London 362~(1821) (2004)
  1655--1665.

\bibitem{MD_Nie_Chen_Robbins}
X.~B. Nie, S.~Y. Chen, W.~E, M.~O. Robbins, A continuum and molecular dynamics
  hybrid method for micro- and nano-fluid flow, Journal of Fluid Mechanics 500
  (2004) 55--64.

\bibitem{MD_Werder_Walther_Koumoutsakos}
T.~Werder, J.~H. Walther, P.~Koumoutsakos, Hybrid atomistic-continuum method
  for the simulation of dense fluid flows, Journal of Computational Physics
  205~(1) (2005) 373--390.

\bibitem{NL_Pennathur_Santiago_2}
S.~Pennathur, J.~G. Santiago, Electrokinetic transport in nanochannels. 2.
  experiments, Analytical Chemistry 77~(21) (2005) 6782--6789.

\bibitem{MD_Rapaport_art}
D.~C. Rapaport, The Art of Molecular Dynamics Simulation, 2nd ed., Cambridge
  University Press, 2004.

\bibitem{MD_Smith_hypercube}
W.~Smith, Molecular dynamics on hypercube parallel computers, Computer Physics
  Communications 62~(2--3) (1991) 229---248.

\bibitem{MD_Rapaport_Million_II}
D.~Rapaport, Multi-million particle molecular dynamics. {II}. {D}esign
  considerations for distributed processing, Computer Physics Communications
  62~(2--3) (1991) 217--228.

\bibitem{MD_Allen_Tildesley}
M.~Allen, D.~Tildesley, Computer simulation of liquids, Oxford University
  Press, 1987.

\bibitem{MP_Weller_Tabor_Jasak_Fureby}
H.~G. Weller, G.~Tabora, H.~Jasak, C.~Fureby, A tensorial approach to
  computational continuum mechanics using object-oriented techniques, Computers
  in Physics 12 (1998) 620--631.

\bibitem{MP_openFOAM}
{OpenFOAM}, \href{http://www.openfoam.org}{www.openfoam.org}.

\bibitem{MD_Allen_Hauser_foamDSMC}
J.~Allen, T.~Hauser, {foamDSMC}: An object oriented parallel {DSMC} solver for
  rarefied flow applications, in: 45th AIAA Aerospace Sciences Meeting and
  Exhibit, Reno, Nevada, 2007.

\end{thebibliography}
\end{document}